\documentclass[useAMS,usenatbib,a4paper]{mnras}
\usepackage{amsmath}
\usepackage{amssymb}
\usepackage{graphicx}
\usepackage{xfrac}
\usepackage{caption}
\usepackage{array}
\usepackage{longtable}
\usepackage{pdflscape}
\usepackage[usenames, dvipsnames]{color}
\usepackage{subcaption}
\usepackage[flushleft]{threeparttable}
\usepackage[export]{adjustbox}
\usepackage{enumitem}
\usepackage{calc}
\newcommand*\vsigma{{v}\hspace{-0.16em}/\hspace{-0.16em}{\sigma}}

\title[SLUGGS trails]{The SLUGGS Survey: Trails of SLUGGS galaxies in a modified spin-ellipticity diagram}
\author[Bellstedt et al.]{Sabine Bellstedt$^{1}$\thanks{Email: sbellstedt@swin.edu.au}, Alister W. Graham$^{1}$, Duncan A. Forbes$^{1}$, \newauthor Aaron J. Romanowsky$^{2, 3}$, Jean P. Brodie$^{3}$, Jay Strader$^{4}$ \\
$^{1}$Centre for Astrophysics and Supercomputing, Swinburne University of Technology, Hawthorn VIC 3122, Australia\\
$^{2}$Department of Physics and Astronomy, San Jos\'{e} State University, One Washington Square, San Jose, CA, 95192, USA\\
$^{3}$University of California Observatories, 1156 High Street, Santa Cruz, CA 95064, USA\\
$^{4}$Department of Physics and Astronomy, Michigan State University, East Lansing, Michigan 48824, USA\\}

\begin{document}

\date{Accepted 2017 May 26. Received 2017 May 21; in original form 2017 April 2}

\pagerange{\pageref{firstpage}--\pageref{lastpage}} \pubyear{2017}

\maketitle

\label{firstpage}

\begin{abstract}

We present radial tracks for four early-type galaxies with embedded intermediate-scale discs in a modified spin-ellipticity diagram. Here, each galaxy's spin and ellipticity profiles are shown as a radial track, as opposed to a single, flux-weighted aperture-dependent value as is common in the literature. 
The use of a single ellipticity and spin parameter is inadequate to capture the basic nature of these galaxies, which transition from fast to slow rotation as one moves to larger radii where the disc ceases to dominate. 
After peaking, the four galaxy's radial tracks feature a downturn in both ellipticity and spin with increasing radius, differentiating them from elliptical galaxies, and from lenticular galaxies whose discs dominate at large radii. These galaxies are examples of so-called discy elliptical galaxies, which are a morphological hybrid between elliptical (E) and lenticular (S0) galaxies, and have been designated ES galaxies. 

The use of spin-ellipticity tracks provides extra structural information about individual galaxies over a single aperture measure. Such tracks provide a key diagnostic for classifying early-type galaxies, particularly in the era of 2D kinematic (and photometric) data beyond one effective radius. 
\end{abstract}


\section{Introduction}

The extent to which rotation is present in galaxies has been a widely explored topic for decades. Rotation profiles for spiral galaxies have been studied since the late 1930s \citep[][]{Babcock39}, but due to the lower surface brightnesses of the outer regions of early-type galaxies (ETGs), studies of their rotation profiles did not commence until later. 
A sample of ETGs were observed to be elongated \citep[][]{Sandage70}, which was expected to be due to rotation \citep[][]{Larson75}. 
Early measurements of rotation profiles in lenticular (S0) galaxies confirmed the expectation of strong rotation \citep[][]{Williams75}.
Astronomers were then surprised by data suggesting that there was little rotation in some elliptical (E) galaxies \citep[e.g.][who measured no rotation in the flattened elliptical NGC 4697]{Bertola75}. 
This unexpected finding led to much research regarding the nature of rotation versus anisotropic velocity dispersion 
\citep[e.g.][]{Illingworth77, Binney78, Schechter79, Davies81, Kormendy82, Davies83}. 

One of the early manners in which information about rotational support was condensed was the so-called \citet{Binney78} diagram. This diagram presented the ratio of the maximum rotational velocity to the central velocity dispersion $v/\sigma$ against the ellipticity $\epsilon$ for individual galaxies. In particular, this diagram was used by \citeauthor{Binney78} to address the question of how galaxies that are not rotationally supported can be elongated by anisotropy. 
\citet{Davies83} later used this diagram and noted that fainter ellipticals ($M_B > -20.5$ mag) tended to be rotationally supported, whereas brighter ellipticals ($M_B < -20.5$ mag) were pressure supported by near-isotropic velocity dispersion.

The use of this diagram was ideal for longslit data, in which a maximum rotation value  along the major axis and a central velocity dispersion could be measured. 
With the increasing use of two-dimensional spectroscopy, in particular via integral field units (IFUs), much more information was available with which to characterise the kinematics of galaxies.  
In order to utilise 2D measurements of both velocity and velocity dispersion within a specified aperture, an observationally-measureable, luminosity-weighted spin parameter, denoted by $\lambda_R$, was developed by the SAURON team \citep{Emsellem07}. 
The use of spatial weighting allowed this parameter to better indicate the rotation of galaxies with nonconventional kinematics (whose properties could not simply be summarised by major-axis measurements, such as galaxies with kinematic twists and misalignments as was later discussed by \citealt{Krajnovic11}). 
The main outcome of the use of the $\lambda_R$ parameter by the SAURON \citep{deZeeuw02} and ATLAS$^{\rm 3D}$ \citep{Cappellari11a} teams was to plot this parameter in conjunction with the ellipticity $\epsilon$ in a spin-ellipticity diagram to identify two classes of early-type galaxies according to their kinematics: fast and slow central rotators. 

Despite the use of high-quality 2D kinematic data, the final results are still typically reduced to a single measurement for each galaxy. 
The breadth of information which can be gained from this diagram is further reduced by the fact that the position in this $\lambda_R\,-\,\epsilon$ space is dependent on the single aperture size, and does not differentiate between galaxies with constant or varying rotation in differing regions within the same galaxy, see figures 2 and 5 in \citet{Emsellem07}. Additionally, ellipticity often changes with radius \citep[as has been noted by numerous studies, e.g.][ and references therein]{Liller66, diTullio78}. 

The use of the $\lambda_R\,-\,\epsilon$ diagram and other kinematic diagnostics has challenged the traditional morphological classification of ETGs into elliptical (E) and lenticular (S0) types, as galaxies previously classified into these morphologies do not form two distinct kinematic populations. 
One reason for this is that inclination affects disc identification differently for photometric and kinematic techniques\footnote{From photometry, discs are identified via their projected ellipticity, which scales with inclination $i$ (where $i=0$ for a face-on galaxy) as $\cos(i)$, whereas kinematic studies identify discs through rotation, which scales as $\sin(i)$.}. 
Moreover, photometric decompositions into bulge and disc components often produce inconsistent results, requiring the inclusion of 2D kinematic data to help break degeneracies in disc fractions and sizes. 
The presence of discs in some ETGs was therefore missed in earlier photometric studies. For example, \citet{Cappellari11b} estimated that only $\sim34$ per cent of elliptical galaxies are correctly classified, and that the rest contain discs. 

The observation of a `morphological hybrid' -- an ETG that displays properties somewhere in between those of ellipticals and lenticulars -- has been noted in the past \citep[][]{Liller66}. The presence of these discs confined within the main spheroidal component of the galaxies has been analysed in photometric studies such as that by \citet{Scorza95}, where the comment was made that there was likely a \textit{continuity of disc properties at the low disc-to-bulge ratio end of the Hubble sequence}. 
Such embedded discs have also been identified by \citet{Cinzano94} and \citet{Rix90}. Recently, \citet{Savorgnan16a} and \citet{Graham16} discussed the presence of `intermediate-scale discs' in ETGs (i.e., discs that are intermediate in size between nuclear discs and large-scale discs), and their downturning ellipticity and spin profiles. Given that such discs come in a range of sizes, \citep{Nieto88, Simien90, Michard93, Andreon96, Krajnovic13} the motivation of separating galaxies purely into the binary classification of fast and slow central rotators could be questioned when considering that galaxies may all have fast and slow rotating components of varying proportions \citep[as is evident in the works of, e.g.][]{Arnold14, Foster16}. 

\citet{Graham17} introduced the concept of plotting galaxies as tracks, rather than as single points, in a modified spin-ellipticity diagram, showing their movement as a function of galaxy radius.
In this paper, we characterize such radial tracks in $\lambda(R)\,-\,\epsilon(R)$ diagrams for galaxies with intermediate-scale discs, and compare these with tracks of typical E and S0 galaxies. In \S 2, we describe our sample and the data used, and in \S 3 we outline our method. The results are given in \S 4, which we discuss in \S 5. We summarise and conclude in \S 6. 

\section{Sample and Data}

\defcitealias{Bellstedt17a}{B17}

We focus on four galaxies from the SLUGGS (SAGES Legacy Unifying Globulars and GalaxieS) survey\footnote{http://sluggs.swin.edu.au} \citep{Brodie14} noted to have declining spin profiles at large radii by \citet[][hereafter \citetalias{Bellstedt17a}]{Bellstedt17a}. 
These galaxies are NGC 821, NGC 3377, NGC 4278 and NGC 4473. All are classified as elliptical galaxies, with circularised effective radii ($R_e$) of 43.2, 45.4, 28.3 and 30.2 arcseconds (4.9, 2.4, 2.1 and 2.2 kpc), respectively.
The kinematics for these galaxies behave differently in their outskirts compared to their inner regions, flagging them as potential examples of galaxies with intermediate-scale discs. NGC 4473 had been described by \citet{Krajnovic11} as a `double-sigma' galaxy, characterised by two velocity dispersion peaks along the major axis, exhibiting two distinct inner and outer rotation components \citep{Foster13, Alabi15}. Hence, NGC 4473 does not host an intermediate-scale disc, however we include it to depict the track diagnostic behaviour for such a galaxy. 
Kinematic maps of all four galaxies using SLUGGS data have been published by \citet{Arnold14} and \citet{Foster16}. Each of these galaxies displays fast rotation in the central region, which declines at larger radii. 

Furthermore, for reference, we present an equivalent analysis for a typical E galaxy (NGC 4365) and a typical S0 galaxy (NGC 1023), and plot radial tracks for other ETGs in the SLUGGS sample. 
The data used within this study all come from the SLUGGS survey, and were taken with the DEIMOS spectrograph on the Keck II telescope in Hawaii. The data reduction procedures are given in previous papers (e.g.\ \citealt{Arnold14}; \citetalias{Bellstedt17a}).

\section{Method}
\label{sec:method}

Rather than simply plotting an aperture $\lambda_R$ value against an average ellipticity, as commonly done in the literature, we plot ``\textit{annular}" values of both stellar spin and ellipticity measured at varying radii to produce tracks for individual galaxies. 
While comparisons of aperture spin measurements of different sizes have been made in the past \citep[see for example, figure 6 in][]{Raskutti14}, our approach provides a clearer understanding of the overall kinematic structure of individual ETGs. 
To do this, we require measurements of both the ellipticity $\epsilon$ and spin $\lambda$ profiles. Ellipticity profiles were taken from \textit{Spitzer} data presented in \citet{Forbes17}, and we made new measurements of $\lambda(R)$ at the corresponding radii from the SLUGGS kinematic data. These ellipticity profiles have been presented in Appendix \ref{sec:appendix} for each of the galaxies for which spin--ellipticity tracks have been plotted.

A slightly modified version of the technique outlined in \citetalias{Bellstedt17a} is used to measure $\lambda(R)$. 
At each radius, an annulus of width $2\arcsec$ is defined with an ellipticity equal to the local isophotal ellipticity of the galaxy, allowing us to measure the local kinematic properties within this annulus. The ellipticity of each annulus varies according to the local measurement, as opposed to a uniform ellipticity value used by \citetalias{Bellstedt17a}. 

As in \citetalias{Bellstedt17a}, 2D maps are produced using the Kriging technique \citep{Pastorello14}. Kriging produces interpolated 2D maps with an effective smoothing scale of $15-25$ arcseconds, depending on the density of data points. 
It is the Kriged map, rather than the sparse data points themselves, that we use to calculate kinematic properties. 
All points in a specified annulus are used to measure both $\lambda(R)$ and $\vsigma(R)$
with the expressions:
$$\lambda(R) \equiv \frac{\sum_{k}R_k|V_k|}{\sum_{k}R_k\sqrt{V_k^2 + \sigma_k^2}}, \,\,\,\,\,{\rm and}\,\,\,\,\,\,\vsigma(R) \equiv \sqrt{\frac{\sum_{k}V_k^2}{\sum_{k}\sigma_k^2}}.$$ 
Here, $R_k$, $V_k$, and $\sigma_k$ represent the circularised radius, velocity and velocity dispersion of the $k$th Kriging point. The differential flux across a single annulus is neglected. As described by \citetalias{Bellstedt17a}, the $\lambda(R)$ measurement in each annulus is taken as the mean value of 100 bootstrap resamplings of the pixels within each annulus. This process ensures that the variation in pixel number per bin (a result of finite pixel resolution) does not affect the measurement. The spread in $\lambda(R)$ values measured in each bin is typically 0.001-0.005, indicating that slight kinematic variations across a single annulus do not affect our measurement of the local stellar spin. 

We find that $\lambda(R)$ and $\vsigma(R)$ scale as
$ \lambda(R) = \kappa\,\vsigma(R)/\sqrt{1 + \kappa^{2}\,\vsigma(R)^2}$
as described by \citet{Emsellem11}, where $\kappa \simeq 0.9$, as opposed to $\kappa \simeq 1.1$ for global values.
The minimum radius at which our tracks begin is $0.1\,R_e$, and the maximum radial extent of our data is defined as the radius where the azimuthal data coverage within the annulus drops to $85$ per cent. 

To determine whether our relatively sparse spatial sampling of the galaxy kinematics, due to our use of multi-slit data rather than a contiguous IFU, affects our results, we compare our tracks with those produced by ATLAS$^{\rm 3D}$ data over the central ($<1\,R_e$) region. We confirm that the results are qualitatively the same. 

\section{Results}

\begin{figure}
	\centering
	\includegraphics[trim = {2mm, 0mm, -2mm, 0mm}, width=90mm]{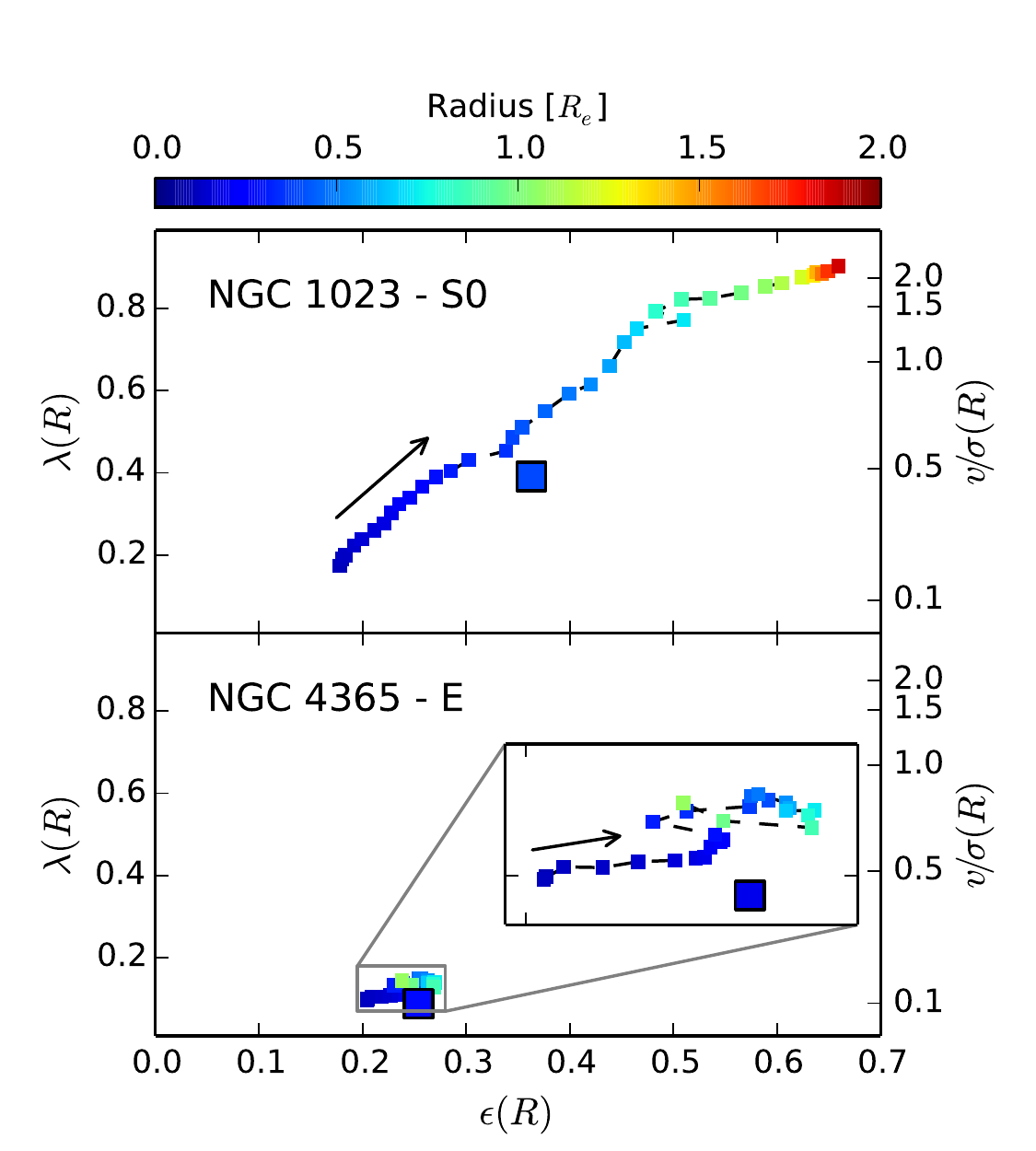}
		\caption{Radial tracks of spin $\lambda(R)$ versus ellipticity $\epsilon(R)$ for a typical S0 galaxy (top panel), and a typical E galaxy (bottom panel). Radius is indicated by the colour of each point. The ATLAS$^{\rm 3D}$ luminosity-weighted value for each galaxy is denoted by the large square. The corresponding $\vsigma(R)$ values, as calculated according to the scaling equation in Section \ref{sec:method}, are indicated by the right axis. Arrows indicate the increasing radial direction for each track. }
		\label{fig:fig1}
 \end{figure}

\begin{figure}
	\centering
	\includegraphics[trim = {2mm, 10mm, -2mm, 0mm}, width=90mm]{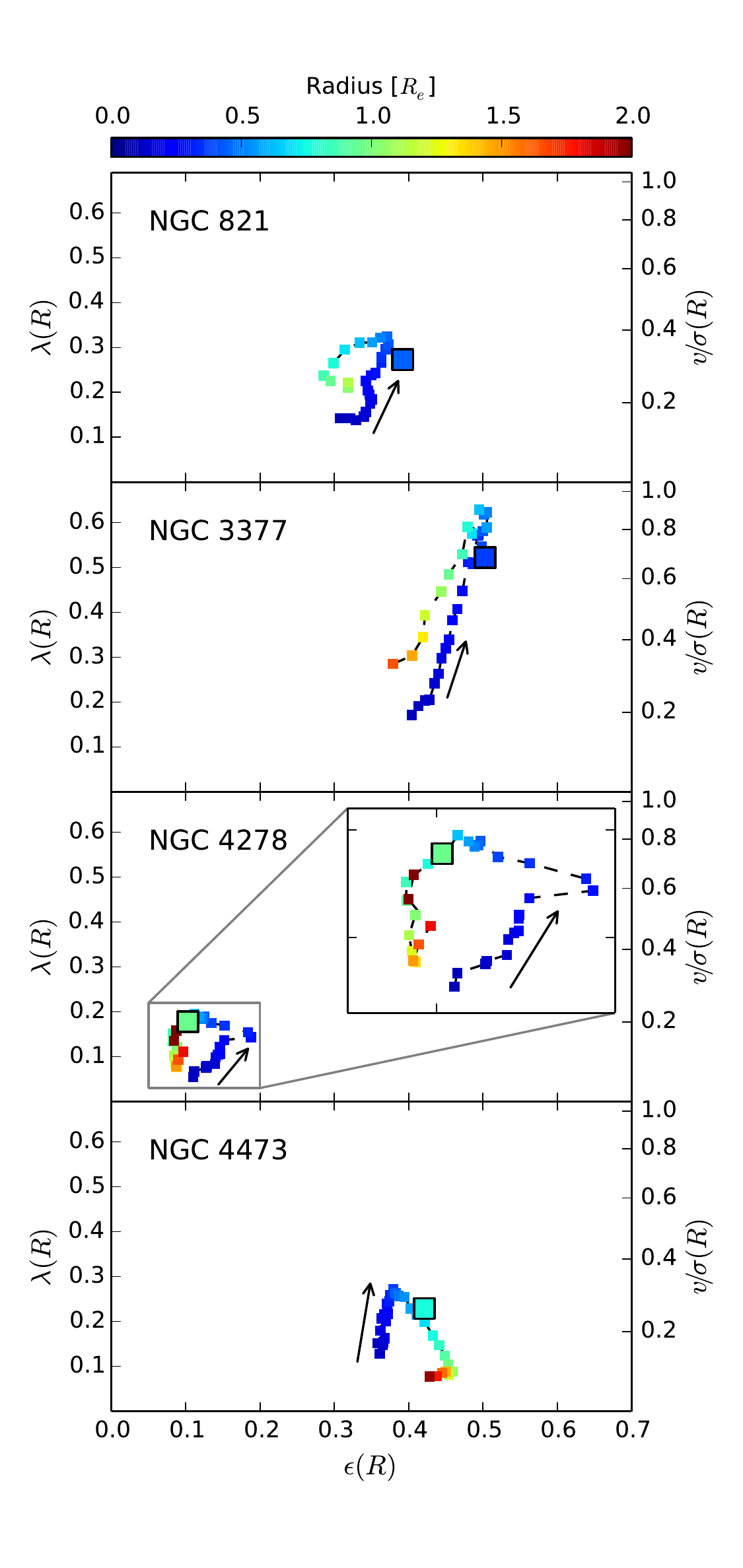}
		\caption{Similar to Figure \ref{fig:fig1} but for four galaxies whose local spin and ellipticity decline at larger radii. Large square points indicate the single aperture measurements from ATLAS$^{\rm 3D}$, whose data extend to 0.44, 0.37, 0.95 and 0.76 $R_e$ for NGC 821, 3377, 4278 and 4473 respectively. }
		\label{fig:fig2}
 \end{figure}

Before presenting the radial tracks of galaxies with intermediate-scale discs, we first display tracks typical of an E and an S0 galaxy. 
Figure \ref{fig:fig1} shows the radial track of the lenticular galaxy NGC 1023, and the slow rotator elliptical galaxy NGC 4365. 
We note that while NGC 4365 has been observed to host a central kinematically distinct core \citep{Krajnovic11}, we do not resolve this feature, and therefore the larger radii kinematics depict those of a typical E galaxy. 

For NGC 1023, the ellipticity increases with radius, becoming more disc-like in its outskirts, as expected for an S0 galaxy containing a large-scale disc that dominates the light at large radii. This increase in ellipticity is associated with a gradual increase in both $\lambda(R)$ and $\vsigma(R)$. 
At larger radii the rotation parameters and ellipticity increase more gradually than in the inner region, indicated by the radial bunching of the data towards the end of the track. Our data, and thus the track for NGC 1023, reaches a radial extent of $\sim2\,R_e$. The `global' $\lambda_R - \epsilon$ value from ATLAS$^{\rm 3D}$ \citep{Emsellem11} is indicated on the plot as a large square. Like the points of the radial track, it is coloured according to the maximum radial extent of the data, i.e. $0.40\,R_e$ (which is why the ATLAS$^{\rm 3D}$ point is shown mid-way along the NGC 1023 radial track). 

The track for NGC 4365 is much more stagnant. The ellipticity is roughly constant at $0.2<\epsilon<0.3$ within $\sim1\,R_e$, and both $\lambda(R)$ and $\vsigma(R)$ show little change with radius. The ATLAS$^{\rm 3D}$ data for this galaxy extend to $0.19\,R_e$. 

Moving on to our four declining spin galaxies, we show the radial tracks of NGC 821, NGC 3377, NGC 4278 and NGC 4473 in Figure \ref{fig:fig2}. For each of these galaxies, it can be seen that there is a downturn with increasing radius in both spin and ellipticity, which generally occurs at $\sim0.5-0.7\,R_e$. When the downturn of both properties occurs at the same radius, the track moves in an anticlockwise direction, as is the case for three out of the four galaxies.
NGC 4473 is slightly different, in that the ellipticity downturn occurs at $\sim1\,R_e$, lagging that of the rotation which occurs at $\sim0.5\,R_e$. This results in a roughly clockwise track. This signature is due to the double-sigma nature of NGC 4473, which has a disc that is counter rotating w.r.t. a slowly-rotating outer region \citep{Krajnovic11}. The counter rotating region has a diluting effect on the disc, which causes the $\lambda(R)$ value to reduce inwards of the radius at which the local ellipticity transitions.

\begin{figure*}
	\centering
	\includegraphics[trim = {0mm, 3mm, 0mm, 2mm}, width=130mm]{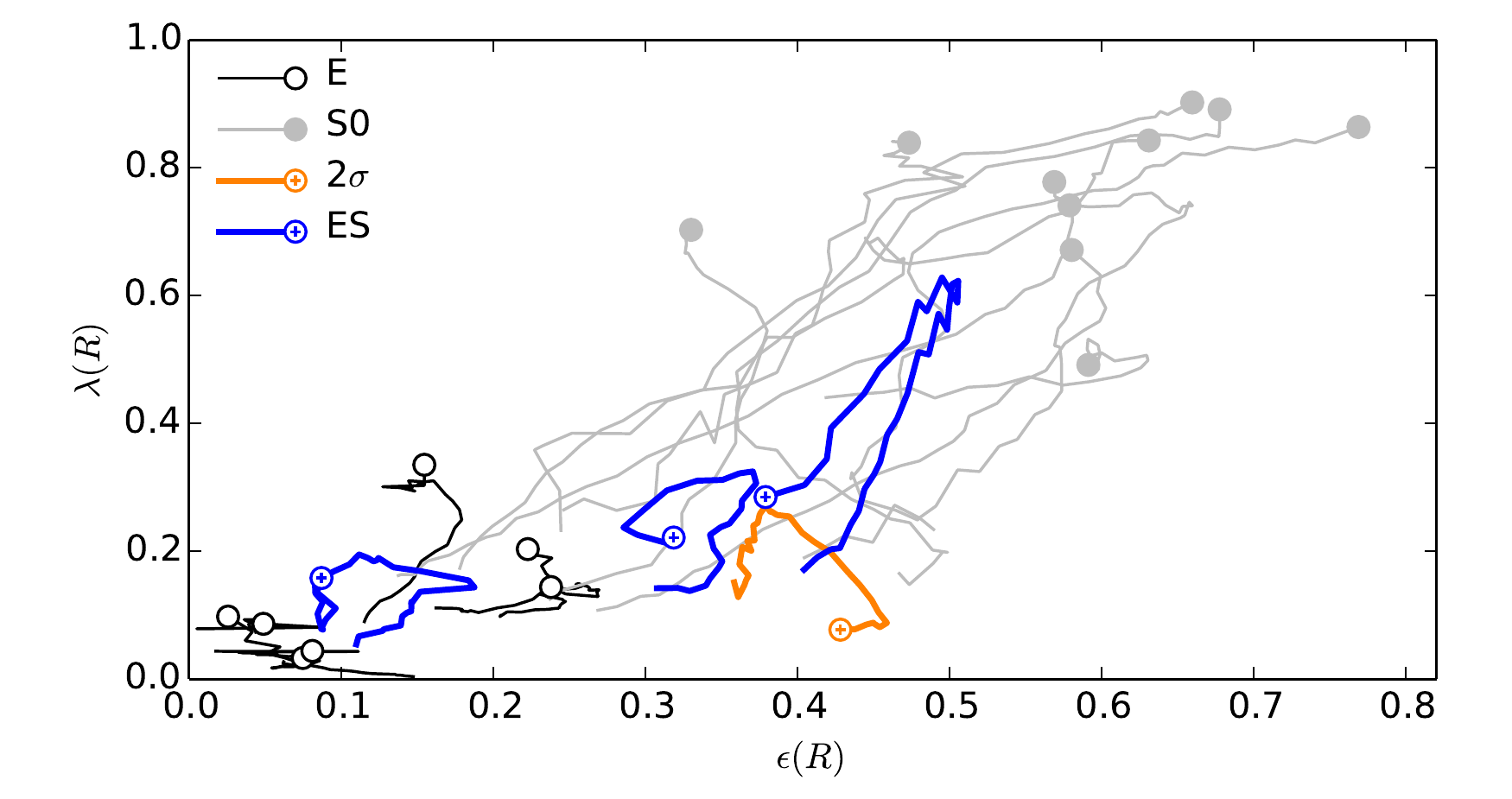}
		\caption{Tracks for SLUGGS galaxies in the modified spin-ellipticity diagram. Markers indicate the maximum radius end of each track. $2\sigma$ refers to the double-sigma galaxy NGC 4473. It can be seen that the structure of the E and ES (NGC 821, 3377, 4278) galaxy tracks are very different. Low-inclination S0 galaxies (see \citetalias{Bellstedt17a} for details) have been omitted from this plot. }
		\label{fig:fig3}
 \end{figure*}

Figure \ref{fig:fig3} displays tracks for 15 additional galaxies from the SLUGGS survey. The data for these tracks have previously been published in \citet{Arnold14}, \citet{Foster16} and \citetalias{Bellstedt17a}. Tracks for S0 galaxies move in a single direction toward the top right region of the plot, whereas tracks for E galaxies hover and remain in the bottom left region of the plot. The tracks for the three hybrid or ES galaxies from Figure \ref{fig:fig2} have been plotted as blue lines, and although the endpoints of the tracks coincide with those of the E galaxies, the structure of the tracks themselves are quite different -- particularly for NGC 3377, which strongly veers into the S0 region before turning down. 
The maximum radius for each track varies between $1-3\,R_e$. 

For clarity, we omit S0 galaxies with somewhat face-on discs \citepalias[as identified in][]{Bellstedt17a} from Figure \ref{fig:fig3}. Going from edge-on ($i=90^{\circ}$) to face-on ($i=0^{\circ}$), the measured rotation reduces from the intrinsic rotational velocity as $v\sin{i}$. The effect of this reduced velocity measurement is to lower the $\lambda(R)$ and $\vsigma(R)$ values. Hence, S0 galaxies with both lower intrinsic rotation and lower inclination will reside in the lower left region of the modified diagram. The same effect is evident in the aperture-based $\lambda_R-\epsilon$ diagram (see figure 15 of \citetalias{Bellstedt17a}).

\section{Discussion}
\label{sec:Discussion}

While the tracks for the three ES galaxies in Figure \ref{fig:fig2} are similar, each is unique in shape and size, revealing valuable insight into their host galaxy's structure and dynamics. Furthermore, these tracks, plus that for the double-sigma galaxy, highlight the amount of information that is not conveyed by single spin and ellipticity values.  
These galaxies display radial tracks very different to those of either typical elliptical or lenticular galaxies. These differences are not seen when summarising the rotational behaviour of galaxies with a single aperture spin and ellipticity value. 
Indeed, all four galaxies are classified as (centrally) fast rotators \citep{Emsellem11}. 

For each galaxy in Figures \ref{fig:fig1} and \ref{fig:fig2}, we plot the $\lambda_{Re}$ value from ATLAS$^{\rm 3D}$ \citep{Emsellem11} as a large square.
When using a larger aperture size, dramatic changes in the spin or ellipticity at greater radii will not strongly influence this value, since the flux-weighted measurement is driven by the rotationally supported galaxy centre \citepalias[as portrayed in figure 9 of][]{Bellstedt17a}. Hence, using larger apertures to measure $\lambda_R$ values will not necessarily be sensistive to the changed behaviour. Since it is this behaviour at greater radii that differentiates the kinematics of galaxies with intermediate-scale discs from those of S0 galaxies, it's capture is important.

Interestingly, NGC 4473 was one of the galaxies classified by \citet{Liller66} as an `ES' galaxy, displaying characteristics midway between an elliptical and a lenticular galaxy, and was referred to as ``ellicular" by \citet{Graham16}. 
NGC 821, despite being previously classified as an E6 galaxy, was determined by \citet{Scorza95} to have a photometric stellar disc, while in the same study NGC 3377 was found to have an intermediate-scale disc within a boxy bulge. This disc within NGC 3377 was also identified by \citet{Arnold14} from kinematics. 
\citet{Raimond81} found an irregular, extended HI disc within NGC 4278 -- a feature unusual in `elliptical' galaxies. 
Hybrid elliptical/lenticular features have therefore been identified previously in each of these galaxies by varying methods, and the $\lambda(R)-\epsilon(R)$ diagram portrays the hybrid nature clearly. 

We expect that the slowly rotating, bulge-like outer regions of galaxies with intermediate-scale discs exhibit tracks similar to the outer regions of elliptical galaxies, with no strong features or variation. Hence, when probed far enough out, the ends of their tracks would likely converge to a fixed point in spin-ellipticity space, unless there is a change of ellipticity in the galaxy outskirts. Although the SLUGGS data have a relatively large radial extent of $\sim2\,R_e$, it is still not far enough to confirm this behaviour for the three galaxies with intermediate-scale discs. Discs can come in a range of sizes relative to the galaxy $R_e$, and as such their decline may or may not be detected by $\sim 2\,R_e$. While NGC 4473 is not a typical spheroidal galaxy with an intermediate-scale disc (it is a double-sigma galaxy), the track does hover around a single point at larger radii, displaying elliptical-type behaviour in its outer regions. 

The presence of intermediate-scale discs has been identified in some local ETGs, which have been interpreted as descendants of high-$z$ compact massive galaxies that have experienced small amounts of disc growth \citep[][]{Graham15, Graham16, Savorgnan16a}. Spheroid masses for NGC 821, 3377 and 4473 were measured by \citet{Savorgnan16}, and determined to range from $3.9 - 4.7\times10^{10}M_{\odot}$. These spheroids are less massive than those of compact massive galaxies ($\sim10^{11}M_{\odot}$) studied by \citet{Graham15}, indicating that embedded, intermediate-scale discs occur within spheroids over a broad stellar mass range. Moreover, \citet{Graham17} report a potential intermediate-scale disc in a dwarf ETG. If one assumes that the structural similarity between ES galaxies is indicative of a common formation path, 
then not only have the most massive `red nuggets' at high-$z$ undergone some degree of disc formation to form present-day galaxies, but this process may apply to high-$z$ compact galaxies with a broad range of stellar masses. 

In order to detect the presence and assess the prevalence of ES galaxies, it is necessary to analyse large galaxy samples. Several IFU galaxy surveys, including SAMI \citep{Bryant15}, MaNGA \citep{Bundy15}, and CALIFA \citep{Sanchez12}, are in varying degrees of completion, and for a subsample of galaxies will reach sufficiently large radii to make such assessments. They will be able to make statistical statements about the prevalence and radial range of intermediate-scale discs in ETGs to build on the statistics provided by the ATLAS$^{\rm 3D}$ survey \citep{Cappellari11a}, and give a clearer indication of the kinematic feature that separates, or more appropriately unites, elliptical and lenticular galaxies. 


\section{Conclusions}

We show that the use of radial tracks in an annulus-based spin-ellipticity $\lambda(R)-\epsilon(R)$ diagram is able to identify a population of  discy elliptical early-type galaxies. This provides a significant advance over existing $\lambda-\epsilon$ diagrams. Discy elliptical galaxies display fast rotation within their inner regions and slow rotation within their outer regions where their discs no longer dominate. In early applications of the spin-ellipticity diagram, which used a single, central flux-weighted value of spin and ellipticity, such galaxies could not be uniquely distinguished from other ETGs and required alternative means of identification. The $\lambda(R)-\epsilon(R)$ diagram provides a succinct method of summarising both kinematic and photometric data within a single diagram, facilitating a quick determination of the galaxy behaviour.


\section{Acknowledgements}

The data presented herein were obtained at the \textit{Spitzer} Space Telescope and the W.M. Keck Observatory. 
We acknowledge the work by Luciana Sinpetru in producing the ellipticity profiles used within this paper.
SB acknowledges the support of the AAO PhD Topup Scholarship. 
AWG and DAF are supported under the ARC's funding schemes DP170102923 and DP130100388 respectively. JPB and AJR are supported by the NSF grants AST-16165598 and AST-1616710 respectively, AJR is supported as a Research Corporation for Science Advancement Cottrell Scholar, and JS is supported by a Packard Fellowship.

\bibliographystyle{mnras}
\setlength{\bibsep}{0.0pt}
\bibliography{BibtexMaster}

\appendix
\section{Ellipticity Profiles}
\label{sec:appendix}

The ellipticity profiles measured from \textit{Spitzer} imaging \citep{Forbes17} are included here for each of the galaxies presented within this paper. The effective radii ($R_e$) for each galaxy can be found in \citet{Forbes17}.

\begin{table}
	\centering
	\setlength\extrarowheight{0pt}
	\caption[NGC821]{NGC 821}
	\label{tab:NGC821Ellipticity}
	\begin{tabular}{@{}|cc|cc|cc|}
		\hline
		Radius & $\epsilon$ & Radius & $\epsilon$ & Radius & $\epsilon$  \\
		($R_e$) &  &  ($R_e$) &  & ($R_e$) &   \\
		\hline
		0.10	&	0.308	 &	0.24	&	0.343	&	0.56	&	0.351 \\
		0.11	&	0.322	 &	0.26	&	0.305	&	0.63	&	0.334 \\
		0.12	&	0.329	 &	0.29	&	0.355	&	0.70	&	0.314 \\
		0.14	&	0.340	 &	0.32	&	0.363	&	0.78	&	0.299 \\
		0.15	&	0.342	 &	0.35	&	0.363	&	0.87	&	0.286 \\
		0.16	&	0.348	 &	0.38	&	0.369	&	0.95	&	0.295 \\
		0.18	&	0.351	 &	0.42	&	0.373	&	1.02	&	0.319 \\
		0.20	&	0.347	 &	0.46	&	0.371	&	1.13	&	0.319 \\
		0.22	&	0.345	 &	0.51	&	0.362	&			&			\\		
		\hline
	\end{tabular}
\end{table}

\begin{table}
	\centering
	\setlength\extrarowheight{0pt}
	\caption[NGC1023]{NGC 1023}
	\label{tab:NGC1023Ellipticity}
	\begin{tabular}{@{}|cc|cc|cc|}
		\hline
		Radius & $\epsilon$ & Radius & $\epsilon$ & Radius & $\epsilon$  \\
($R_e$) &  &  ($R_e$) &  & ($R_e$) &   \\
		\hline
0.10	&	0.178	&	0.30	&	0.285	&	0.80	&	0.483	\\
0.11	&	0.180	&	0.33	&	0.303	&	0.86	&	0.508	\\
0.12	&	0.183	&	0.35	&	0.339	&	0.92	&	0.535	\\
0.14	&	0.192	&	0.38	&	0.345	&	0.98	&	0.566	\\
0.15	&	0.199	&	0.42	&	0.354	&	1.05	&	0.589	\\
0.16	&	0.211	&	0.45	&	0.376	&	1.13	&	0.605	\\
0.18	&	0.221	&	0.49	&	0.399	&	1.21	&	0.624	\\
0.19	&	0.228	&	0.53	&	0.420	&	1.31	&	0.635	\\
0.21	&	0.235	&	0.57	&	0.439	&	1.44	&	0.638	\\
0.23	&	0.246	&	0.62	&	0.453	&	1.57	&	0.643	\\
0.25	&	0.258	&	0.68	&	0.465	&	1.72	&	0.649	\\
0.28	&	0.271	&	0.71	&	0.510	&	1.86	&	0.659	\\
		\hline
\end{tabular}
\end{table}

\begin{table}
	\centering
	\setlength\extrarowheight{0pt}
	\caption[NGC1407]{NGC 1407}
	\label{tab:NGC1407Ellipticity}
	\begin{tabular}{@{}|cc|cc|cc|}
		\hline
		Radius & $\epsilon$ & Radius & $\epsilon$ & Radius & $\epsilon$  \\
($R_e$) &  &  ($R_e$) &  & ($R_e$) &   \\
		\hline
0.10	&	0.044	 &	0.22	&	0.044	&	0.42	&	0.036	 \\
0.11	&	0.044	 &	0.24	&	0.047	&	0.46	&	0.044	 \\
0.12	&	0.041	 &	0.26	&	0.048	&	0.51	&	0.061	 \\
0.13	&	0.039	 &	0.29	&	0.046	&	0.56	&	0.042	 \\
0.15	&	0.040	 &	0.32	&	0.053	&	0.63	&	0.005	 \\
0.16	&	0.042	 &	0.35	&	0.053	&	0.66	&	0.087	 \\
0.18	&	0.048	 &	0.38	&	0.048	&	0.74	&	0.049	 \\
0.20	&	0.044	 &			&			&			&			\\
		\hline
	\end{tabular}
\end{table}

\begin{table}
	\centering
	\setlength\extrarowheight{0pt}
	\caption[NGC2549]{NGC 2549}
	\label{tab:NGC2549Ellipticity}
	\begin{tabular}{@{}|cc|cc|cc|}
		\hline
		Radius & $\epsilon$ & Radius & $\epsilon$ & Radius & $\epsilon$  \\
($R_e$) &  &  ($R_e$) &  & ($R_e$) &   \\
		\hline
0.10	&	0.244	 &	0.30	&	0.442	 &	0.98	&	0.510	 \\
0.11	&	0.244	 &	0.33	&	0.463	 &	1.07	&	0.525	 \\
0.12	&	0.234	 &	0.36	&	0.469	 &	1.15	&	0.544	 \\
0.14	&	0.227	 &	0.39	&	0.470	 &	1.23	&	0.565	 \\
0.15	&	0.233	 &	0.44	&	0.467	 &	1.33	&	0.581	 \\
0.16	&	0.252	 &	0.48	&	0.459	 &	1.45	&	0.590	 \\
0.18	&	0.290	 &	0.53	&	0.453	 &	1.58	&	0.595	 \\
0.19	&	0.314	 &	0.59	&	0.447	 &	1.74	&	0.594	 \\
0.21	&	0.337	 &	0.65	&	0.445	 &	1.91	&	0.600	 \\
0.22	&	0.362	 &	0.71	&	0.450	 &	2.08	&	0.607	 \\
0.24	&	0.380	 &	0.78	&	0.457	 &	2.26	&	0.615	 \\
0.26	&	0.396	 &	0.84	&	0.474	 &	2.44	&	0.631	 \\
0.28	&	0.418	 &	0.91	&	0.497	 &			&			 \\
		\hline
	\end{tabular}
\end{table}

\begin{table}
	\centering
	\setlength\extrarowheight{0pt}
	\caption[NGC2768]{NGC 2768}
	\label{tab:NGC2768Ellipticity}
	\begin{tabular}{@{}|cc|cc|cc|}
		\hline
		Radius & $\epsilon$ & Radius & $\epsilon$ & Radius & $\epsilon$  \\
($R_e$) &  &  ($R_e$) &  & ($R_e$) &   \\
		\hline
0.10	&	0.269	 &	0.26	&	0.406	 &	0.67	&	0.531	 \\ 
0.11	&	0.281	 &	0.29	&	0.421	 &	0.73	&	0.542	 \\ 
0.12	&	0.297	 &	0.31	&	0.436	 &	0.79	&	0.553	 \\ 
0.13	&	0.309	 &	0.34	&	0.450	 &	0.86	&	0.563	 \\ 
0.14	&	0.323	 &	0.37	&	0.468	 &	0.93	&	0.576	 \\ 
0.16	&	0.332	 &	0.40	&	0.480	 &	1.01	&	0.588	 \\ 
0.17	&	0.342	 &	0.43	&	0.493	 &	1.10	&	0.599	 \\ 
0.19	&	0.353	 &	0.47	&	0.502	 &	1.20	&	0.603	 \\ 
0.21	&	0.364	 &	0.51	&	0.510	 &	1.33	&	0.597	 \\ 
0.22	&	0.377	 &	0.56	&	0.513	 &	1.46	&	0.599	 \\ 
0.24	&	0.390	 &	0.61	&	0.522	 &	1.64	&	0.580	 \\ 		
		\hline
	\end{tabular}
\end{table}

\begin{table}
	\centering
	\setlength\extrarowheight{0pt}
	\caption[NGC2974]{NGC 2974}
	\label{tab:NGC2974Ellipticity}
	\begin{tabular}{@{}|cc|cc|cc|}
		\hline
		Radius & $\epsilon$ & Radius & $\epsilon$ & Radius & $\epsilon$  \\
($R_e$) &  &  ($R_e$) &  & ($R_e$) &   \\
		\hline
0.11	&	0.237	 &	 0.24	&	0.329	 &	 0.54	&	0.372	 \\ 
0.12	&	0.245	 &	 0.26	&	0.344	 &	 0.60	&	0.354	 \\ 
0.13	&	0.259	 &	 0.28	&	0.352	 &	 0.67	&	0.338	 \\ 
0.14	&	0.281	 &	 0.31	&	0.359	 &	 0.74	&	0.334	 \\ 
0.15	&	0.304	 &	 0.34	&	0.359	 &	 0.82	&	0.328	 \\ 
0.16	&	0.309	 &	 0.37	&	0.363	 &	 0.90	&	0.326	 \\ 
0.18	&	0.315	 &	 0.41	&	0.370	 &	 0.99	&	0.327	 \\ 
0.20	&	0.324	 &	 0.45	&	0.376	 &	 1.09	&	0.326	 \\ 
0.22	&	0.324	 &	 0.49	&	0.380	 &	 1.20	&	0.330	 \\ 
		\hline
	\end{tabular}
\end{table}

\begin{table}
	\centering
	\setlength\extrarowheight{0pt}
	\caption[NGC3115]{NGC 3115}
	\label{tab:NGC3115Ellipticity}
	\begin{tabular}{@{}|cc|cc|cc|}
		\hline
		Radius & $\epsilon$ & Radius & $\epsilon$ & Radius & $\epsilon$  \\
($R_e$) &  &  ($R_e$) &  & ($R_e$) &   \\
		\hline
0.10	&	0.405	 &	0.31	&	0.574	 &	0.95	&	0.652	 \\ 
0.11	&	0.413	 &	0.34	&	0.574	 &	1.04	&	0.655	 \\ 
0.12	&	0.431	 &	0.37	&	0.573	 &	1.14	&	0.657	 \\ 
0.13	&	0.450	 &	0.41	&	0.569	 &	1.25	&	0.660	 \\ 
0.15	&	0.464	 &	0.45	&	0.571	 &	1.39	&	0.653	 \\ 
0.16	&	0.480	 &	0.49	&	0.576	 &	1.54	&	0.651	 \\ 
0.17	&	0.491	 &	0.53	&	0.584	 &	1.71	&	0.642	 \\ 
0.19	&	0.505	 &	0.58	&	0.592	 &	1.90	&	0.637	 \\ 
0.20	&	0.520	 &	0.63	&	0.606	 &	2.10	&	0.633	 \\ 
0.22	&	0.532	 &	0.68	&	0.615	 &	2.34	&	0.621	 \\ 
0.24	&	0.544	 &	0.74	&	0.623	 &	2.61	&	0.611	 \\ 
0.26	&	0.556	 &	0.81	&	0.631	 &	2.94	&	0.592	 \\ 
0.28	&	0.565	 &	0.88	&	0.640	 &	3.29	&	0.579	 \\ 		
		\hline
	\end{tabular}
\end{table}

\begin{table}
	\centering
	\setlength\extrarowheight{0pt}
	\caption[NGC3608]{NGC 3608}
	\label{tab:NGC3608Ellipticity}
	\begin{tabular}{@{}|cc|cc|cc|}
		\hline
		Radius & $\epsilon$ & Radius & $\epsilon$ & Radius & $\epsilon$  \\
($R_e$) &  &  ($R_e$) &  & ($R_e$) &   \\
		\hline
0.11	&	0.162	 &	 0.25	&	0.179	 &	0.56	&	0.232	 \\ 
0.12	&	0.165	 &	 0.27	&	0.179	 &	0.61	&	0.240	 \\ 
0.13	&	0.171	 &	 0.30	&	0.181	 &	0.68	&	0.235	 \\ 
0.14	&	0.173	 &	 0.33	&	0.183	 &	0.75	&	0.232	 \\ 
0.15	&	0.177	 &	 0.36	&	0.185	 &	0.82	&	0.239	 \\ 
0.17	&	0.178	 &	 0.39	&	0.190	 &	0.91	&	0.226	 \\ 
0.18	&	0.180	 &	 0.43	&	0.203	 &	1.00	&	0.220	 \\ 
0.20	&	0.180	 &	 0.47	&	0.214	 &	1.10	&	0.223	 \\ 
0.22	&	0.179	 &	 0.51	&	0.226	 &			&			 \\		
		\hline
	\end{tabular}
\end{table}

\begin{table}
	\centering
	\setlength\extrarowheight{0pt}
	\caption[NGC3377]{NGC 3377}
	\label{tab:NGC3377Ellipticity}
	\begin{tabular}{@{}|cc|cc|cc|}
		\hline
		Radius & $\epsilon$ & Radius & $\epsilon$ & Radius & $\epsilon$  \\
($R_e$) &  &  ($R_e$) &  & ($R_e$) &   \\
		\hline
0.10	&	0.404	&	0.25	&	0.466	&	 0.63	&	0.495	 \\
0.11	&	0.413	&	0.27	&	0.472	&	 0.70	&	0.485	 \\
0.12	&	0.422	&	0.30	&	0.480	&	 0.77	&	0.479	 \\
0.13	&	0.428	&	0.33	&	0.486	&	 0.86	&	0.472	 \\
0.15	&	0.435	&	0.36	&	0.493	&	 0.96	&	0.454	 \\
0.16	&	0.440	&	0.39	&	0.498	&	 1.07	&	0.444	 \\
0.17	&	0.444	&	0.43	&	0.500	&	 1.19	&	0.422	 \\
0.19	&	0.450	&	0.47	&	0.502	&	 1.32	&	0.419	 \\
0.21	&	0.454	&	0.52	&	0.506	&	 1.47	&	0.405	 \\
0.23	&	0.459	&	0.57	&	0.505	&	 1.65	&	0.379	 \\		
		\hline
	\end{tabular}
\end{table}

\begin{table}
	\centering
	\setlength\extrarowheight{0pt}
	\caption[NGC4111]{NGC 4111}
	\label{tab:NGC4111Ellipticity}
	\begin{tabular}{@{}|cc|cc|cc|}
		\hline
		Radius & $\epsilon$ & Radius & $\epsilon$ & Radius & $\epsilon$  \\
($R_e$) &  &  ($R_e$) &  & ($R_e$) &   \\
		\hline
0.10	&	0.489	 &	0.42	&	0.499	 &	1.39	&	0.618	 \\ 
0.11	&	0.480	 &	0.46	&	0.495	 &	1.52	&	0.624	 \\ 
0.12	&	0.479	 &	0.52	&	0.479	 &	1.66	&	0.629	 \\ 
0.14	&	0.465	 &	0.57	&	0.473	 &	1.81	&	0.634	 \\ 
0.16	&	0.443	 &	0.63	&	0.476	 &	1.97	&	0.641	 \\ 
0.17	&	0.436	 &	0.69	&	0.482	 &	2.13	&	0.653	 \\ 
0.19	&	0.435	 &	0.75	&	0.493	 &	2.30	&	0.668	 \\ 
0.21	&	0.442	 &	0.81	&	0.507	 &	2.47	&	0.684	 \\ 
0.23	&	0.447	 &	0.88	&	0.523	 &	2.64	&	0.701	 \\ 
0.24	&	0.465	 &	0.95	&	0.541	 &	2.82	&	0.717	 \\ 
0.27	&	0.466	 &	1.02	&	0.561	 &	3.05	&	0.727	 \\ 
0.29	&	0.469	 &	1.10	&	0.578	 &	3.27	&	0.740	 \\ 
0.33	&	0.467	 &	1.18	&	0.594	 &	3.53	&	0.750	 \\ 
0.36	&	0.469	 &	1.28	&	0.610	 &	3.74	&	0.769	 \\ 
0.39	&	0.487	 &			&			&			&			 \\		
		\hline
	\end{tabular}
\end{table}

\begin{table}
	\centering
	\setlength\extrarowheight{0pt}
	\caption[NGC4278]{NGC 4278}
	\label{tab:NGC4278Ellipticity}
	\begin{tabular}{@{}|cc|cc|cc|}
		\hline
		Radius & $\epsilon$ & Radius & $\epsilon$ & Radius & $\epsilon$  \\
($R_e$) &  &  ($R_e$) &  & ($R_e$) &   \\
		\hline
0.10	&	0.110	&	0.28	&	0.188	&	0.85	&	0.083	 \\
0.11	&	0.112	&	0.31	&	0.184	&	0.93	&	0.083	 \\
0.12	&	0.127	&	0.34	&	0.152	&	1.02	&	0.088	 \\
0.13	&	0.128	&	0.38	&	0.134	&	1.12	&	0.085	 \\
0.15	&	0.139	&	0.42	&	0.125	&	1.24	&	0.086	 \\
0.16	&	0.140	&	0.47	&	0.123	&	1.36	&	0.088	 \\
0.18	&	0.143	&	0.51	&	0.121	&	1.50	&	0.087	 \\
0.20	&	0.146	&	0.57	&	0.118	&	1.64	&	0.090	 \\
0.22	&	0.146	&	0.63	&	0.112	&	1.80	&	0.097	 \\
0.24	&	0.146	&	0.69	&	0.105	&	1.99	&	0.084	 \\
0.26	&	0.152	&	0.76	&	0.095	&	2.19	&	0.087	 \\		
		\hline
	\end{tabular}
\end{table}

\clearpage

\begin{table}
	\centering
	\setlength\extrarowheight{0pt}
	\caption[NGC4365]{NGC 4365}
	\label{tab:NGC4365Ellipticity}
	\begin{tabular}{@{}|cc|cc|cc|}
		\hline
		Radius & $\epsilon$ & Radius & $\epsilon$ & Radius & $\epsilon$  \\
($R_e$) &  &  ($R_e$) &  & ($R_e$) &   \\
		\hline
		0.10	&	0.205	&	0.23	&	0.248	&	0.54	&	0.263	\\
		0.11	&	0.204	&	0.25	&	0.247	&	0.59	&	0.264	\\
		0.12	&	0.209	&	0.28	&	0.246	&	0.65	&	0.263	\\
		0.13	&	0.218	&	0.31	&	0.231	&	0.71	&	0.27	\\
		0.14	&	0.227	&	0.34	&	0.239	&	0.78	&	0.268	\\
		0.16	&	0.236	&	0.37	&	0.254	&	0.86	&	0.269	\\
		0.17	&	0.241	&	0.41	&	0.259	&	0.96	&	0.248	\\
		0.19	&	0.243	&	0.45	&	0.254	&	1.07	&	0.238	\\
		0.21	&	0.244	&	0.49	&	0.256	&			&			\\
		\hline
	\end{tabular}
\end{table}

\begin{table}
	\centering
	\setlength\extrarowheight{0pt}
	\caption[NGC4374]{NGC 4374}
	\label{tab:NGC4374Ellipticity}
	\begin{tabular}{@{}|cc|cc|cc|}
		\hline
		Radius & $\epsilon$ & Radius & $\epsilon$ & Radius & $\epsilon$  \\
($R_e$) &  &  ($R_e$) &  & ($R_e$) &   \\
		\hline
0.10	&	0.147	 &	0.23	&	0.114	 &	0.45	&	0.067	 \\ 
0.11	&	0.144	 &	0.25	&	0.109	 &	0.50	&	0.063	 \\ 
0.13	&	0.141	 &	0.27	&	0.104	 &	0.55	&	0.051	 \\ 
0.14	&	0.141	 &	0.30	&	0.093	 &	0.60	&	0.060	 \\ 
0.15	&	0.136	 &	0.33	&	0.089	 &	0.67	&	0.037	 \\ 
0.17	&	0.136	 &	0.37	&	0.080	 &	0.74	&	0.042	 \\ 
0.18	&	0.126	 &	0.41	&	0.079	 &	0.82	&	0.026	 \\ 
0.20	&	0.120	 &			&			 &			&				\\		
		\hline
	\end{tabular}
\end{table}

\begin{table}
	\centering
	\setlength\extrarowheight{0pt}
	\caption[NGC4473]{NGC 4473}
	\label{tab:NGC4473Ellipticity}
	\begin{tabular}{@{}|cc|cc|cc|}
		\hline
		Radius & $\epsilon$ & Radius & $\epsilon$ & Radius & $\epsilon$  \\
($R_e$) &  &  ($R_e$) &  & ($R_e$) &   \\
		\hline
0.11	&	0.358	&	0.31	&	0.374	&	0.82	&	0.441	 \\
0.12	&	0.361	&	0.34	&	0.375	&	0.90	&	0.449	 \\
0.13	&	0.365	&	0.37	&	0.379	&	0.99	&	0.453	 \\
0.14	&	0.366	&	0.40	&	0.382	&	1.08	&	0.459	 \\
0.16	&	0.368	&	0.44	&	0.382	&	1.19	&	0.454	 \\
0.17	&	0.362	&	0.49	&	0.387	&	1.32	&	0.450	 \\
0.19	&	0.364	&	0.53	&	0.395	&	1.45	&	0.450	 \\
0.21	&	0.370	&	0.58	&	0.403	&	1.60	&	0.445	 \\
0.23	&	0.367	&	0.64	&	0.411	&	1.77	&	0.438	 \\
0.25	&	0.372	&	0.69	&	0.422	&	1.96	&	0.428	 \\
0.28	&	0.371	&	0.75	&	0.433	&	2.17	&	0.422	 \\		
		\hline
	\end{tabular}
\end{table}

\begin{table}
	\centering
	\setlength\extrarowheight{0pt}
	\caption[NGC4486]{NGC 4486}
	\label{tab:NGC4486Ellipticity}
	\begin{tabular}{@{}|cc|cc|cc|}
		\hline
		Radius & $\epsilon$ & Radius & $\epsilon$ & Radius & $\epsilon$  \\
($R_e$) &  &  ($R_e$) &  & ($R_e$) &   \\
		\hline
0.10	&	0.018	 &	0.23	&	0.054	 &	0.50	&	0.044	 \\ 
0.11	&	0.021	 &	0.26	&	0.047	 &	0.55	&	0.044	 \\ 
0.12	&	0.031	 &	0.28	&	0.049	 &	0.60	&	0.059	 \\ 
0.13	&	0.062	 &	0.31	&	0.037	 &	0.66	&	0.057	 \\ 
0.14	&	0.094	 &	0.34	&	0.033	 &	0.72	&	0.065	 \\ 
0.15	&	0.111	 &	0.38	&	0.032	 &	0.79	&	0.069	 \\ 
0.17	&	0.090	 &	0.42	&	0.035	 &	0.87	&	0.072	 \\ 
0.19	&	0.075	 &	0.46	&	0.039	 &	0.96	&	0.081	 \\ 
0.21	&	0.077	 &			& 			 &			&			 \\		
		\hline
	\end{tabular}
\end{table}

\begin{table}
	\centering
	\setlength\extrarowheight{0pt}
	\caption[NGC4494]{NGC 4494}
	\label{tab:NGC4494Ellipticity}
	\begin{tabular}{@{}|cc|cc|cc|}
		\hline
		Radius & $\epsilon$ & Radius & $\epsilon$ & Radius & $\epsilon$  \\
($R_e$) &  &  ($R_e$) &  & ($R_e$) &   \\
		\hline
0.11	&	0.115	 &	0.27	&	0.153	 &	0.71	&	0.150	 \\ 
0.12	&	0.116	 &	0.30	&	0.162	 &	0.78	&	0.155	 \\ 
0.13	&	0.117	 &	0.33	&	0.170	 &	0.86	&	0.144	 \\ 
0.14	&	0.119	 &	0.36	&	0.175	 &	0.94	&	0.147	 \\ 
0.16	&	0.126	 &	0.39	&	0.179	 &	1.04	&	0.143	 \\ 
0.17	&	0.133	 &	0.43	&	0.178	 &	1.14	&	0.149	 \\ 
0.19	&	0.138	 &	0.48	&	0.173	 &	1.26	&	0.139	 \\ 
0.21	&	0.145	 &	0.53	&	0.169	 &	1.40	&	0.127	 \\ 
0.23	&	0.149	 &	0.58	&	0.161	 &	1.51	&	0.155	 \\ 
0.25	&	0.150	 &	0.64	&	0.152	 &	1.66	&	0.155	 \\ 		
		\hline
	\end{tabular}
\end{table}

\begin{table}
	\centering
	\setlength\extrarowheight{0pt}
	\caption[NGC4526]{NGC 4526}
	\label{tab:NGC4526Ellipticity}
	\begin{tabular}{@{}|cc|cc|cc|}
		\hline
		Radius & $\epsilon$ & Radius & $\epsilon$ & Radius & $\epsilon$  \\
($R_e$) &  &  ($R_e$) &  & ($R_e$) &   \\
		\hline
0.10	&	0.467	 &	0.38	&	0.373	 &	1.16	&	0.504	 \\ 
0.11	&	0.473	 &	0.42	&	0.361	 &	1.26	&	0.524	 \\ 
0.12	&	0.489	 &	0.46	&	0.360	 &	1.35	&	0.545	 \\ 
0.13	&	0.496	 &	0.51	&	0.362	 &	1.44	&	0.569	 \\ 
0.14	&	0.499	 &	0.56	&	0.360	 &	1.54	&	0.596	 \\ 
0.16	&	0.489	 &	0.61	&	0.372	 &	1.64	&	0.619	 \\ 
0.18	&	0.474	 &	0.66	&	0.387	 &	1.78	&	0.632	 \\ 
0.20	&	0.461	 &	0.72	&	0.407	 &	1.92	&	0.647	 \\ 
0.22	&	0.450	 &	0.78	&	0.427	 &	2.07	&	0.658	 \\ 
0.25	&	0.435	 &	0.84	&	0.447	 &	2.24	&	0.669	 \\ 
0.27	&	0.418	 &	0.91	&	0.461	 &	2.44	&	0.677	 \\ 
0.31	&	0.401	 &	0.99	&	0.474	 &	2.68	&	0.677	 \\ 
0.34	&	0.386	 &	1.08	&	0.488	 &	2.95	&	0.677	 \\ 		
		\hline
	\end{tabular}
\end{table}

\begin{table}
	\centering
	\setlength\extrarowheight{0pt}
	\caption[NGC4564]{NGC 4564}
	\label{tab:NGC4564Ellipticity}
	\begin{tabular}{@{}|cc|cc|cc|}
		\hline
		Radius & $\epsilon$ & Radius & $\epsilon$ & Radius & $\epsilon$  \\
($R_e$) &  &  ($R_e$) &  & ($R_e$) &   \\
		\hline
0.11	&	0.138	 &	0.34	&	0.281	 &	0.86	&	0.536	 \\
0.12	&	0.140	 &	0.37	&	0.301	 &	0.94	&	0.549	 \\
0.13	&	0.148	 &	0.40	&	0.322	 &	1.02	&	0.560	 \\
0.14	&	0.153	 &	0.44	&	0.345	 &	1.11	&	0.568	 \\
0.16	&	0.157	 &	0.47	&	0.368	 &	1.21	&	0.573	 \\
0.17	&	0.171	 &	0.51	&	0.390	 &	1.33	&	0.577	 \\
0.19	&	0.183	 &	0.55	&	0.414	 &	1.46	&	0.578	 \\
0.20	&	0.196	 &	0.59	&	0.438	 &	1.60	&	0.581	 \\
0.22	&	0.205	 &	0.64	&	0.463	 &	1.76	&	0.581	 \\
0.24	&	0.215	 &	0.68	&	0.486	 &	1.94	&	0.578	 \\
0.27	&	0.230	 &	0.74	&	0.504	 &	2.16	&	0.569	 \\
0.29	&	0.243	 &	0.80	&	0.523	 &	2.38	&	0.569	 \\
0.32	&	0.261	 &			&			 &			&			 \\		
		\hline
	\end{tabular}
\end{table}

\begin{table}
	\centering
	\setlength\extrarowheight{0pt}
	\caption[NGC5846]{NGC 5846}
	\label{tab:NGC5846Ellipticity}
	\begin{tabular}{@{}|cc|cc|cc|}
		\hline
		Radius & $\epsilon$ & Radius & $\epsilon$ & Radius & $\epsilon$  \\
($R_e$) &  &  ($R_e$) &  & ($R_e$) &   \\
		\hline
0.10	&	0.062	 &	0.20	&	0.059	 &	0.39	&	0.062	 \\ 
0.11	&	0.062	 &	0.22	&	0.057	 &	0.43	&	0.067	 \\ 
0.13	&	0.062	 &	0.25	&	0.055	 &	0.47	&	0.079	 \\ 
0.14	&	0.064	 &	0.27	&	0.054	 &	0.52	&	0.078	 \\ 
0.15	&	0.068	 &	0.30	&	0.059	 &	0.57	&	0.082	 \\ 
0.17	&	0.071	 &	0.33	&	0.058	 &	0.63	&	0.086	 \\ 
0.18	&	0.062	 &	0.36	&	0.061	 &	0.69	&	0.075	 \\ 		
		\hline
	\end{tabular}
\end{table}

\begin{table}
	\centering
	\setlength\extrarowheight{0pt}
	\caption[NGC5866]{NGC 5866}
	\label{tab:NGC5866Ellipticity}
	\begin{tabular}{@{}|cc|cc|cc|}
		\hline
		Radius & $\epsilon$ & Radius & $\epsilon$ & Radius & $\epsilon$  \\
($R_e$) &  &  ($R_e$) &  & ($R_e$) &   \\
		\hline
		0.21	&	0.419	 &	0.42	&	0.592	 &	1.00	&	0.593	 \\ 
		0.23	&	0.439	 &	0.45	&	0.613	 &	1.10	&	0.590	 \\ 
		0.25	&	0.457	 &	0.49	&	0.623	 &	1.21	&	0.591	 \\ 
		0.27	&	0.474	 &	0.54	&	0.626	 &	1.32	&	0.598	 \\ 
		0.29	&	0.490	 &	0.59	&	0.630	 &	1.45	&	0.599	 \\ 
		0.32	&	0.512	 &	0.65	&	0.630	 &	1.60	&	0.598	 \\ 
		0.34	&	0.531	 &	0.72	&	0.623	 &	1.76	&	0.598	 \\ 
		0.37	&	0.552	 &	0.81	&	0.609	 &	1.94	&	0.593	 \\ 
		0.39	&	0.575	 &	0.90	&	0.600  	 &	2.14	&	0.591	 \\ 		
		\hline
	\end{tabular}
\end{table}

\begin{table}
	\centering
	\setlength\extrarowheight{0pt}
	\caption[NGC7457]{NGC 7457}
	\label{tab:NGC7457Ellipticity}
	\begin{tabular}{@{}|cc|cc|cc|}
		\hline
		Radius & $\epsilon$ & Radius & $\epsilon$ & Radius & $\epsilon$  \\
($R_e$) &  &  ($R_e$) &  & ($R_e$) &   \\
		\hline
0.10	&	0.246	 &	0.27	&	0.399	 &	0.65	&	0.471	 \\ 
0.11	&	0.260	 &	0.29	&	0.410	 &	0.68	&	0.509	 \\ 
0.12	&	0.281	 &	0.32	&	0.413	 &	0.77	&	0.481	 \\ 
0.13	&	0.306	 &	0.35	&	0.417	 &	0.86	&	0.467	 \\ 
0.15	&	0.308	 &	0.38	&	0.422	 &	0.94	&	0.473	 \\ 
0.16	&	0.316	 &	0.42	&	0.420	 &	1.05	&	0.457	 \\ 
0.17	&	0.336	 &	0.46	&	0.421	 &	1.15	&	0.464	 \\ 
0.19	&	0.345	 &	0.50	&	0.444	 &	1.25	&	0.477	 \\ 
0.21	&	0.352	 &	0.54	&	0.447	 &	1.39	&	0.462	 \\ 
0.22	&	0.371	 &	0.60	&	0.444	 &	1.52	&	0.473	 \\ 
0.24	&	0.387	 &			& 			 &  		&   		\\		
		\hline
	\end{tabular}
\end{table}

\end{document}